\documentclass{elsart}
\usepackage{natbib}
\usepackage{epsfig}

\begin{document}

%
%
%
\setcounter{page}{1}
\begin{frontmatter}
\title{An electromagnetic shashlik calorimeter
with longitudinal segmentation}
\author[bologna]{A.C.~Benvenuti},
\author[ihep]{I.~Britvich},
\author[cern]{T~.Camporesi}, 
\author[padua]{P.~Checchia}, 
\author[ihep]{A.~Feniouk}, 
\author[lund]{V.~Hedberg}, 
\author[ihep]{V.~Lishin}, 
\author[padua]{M.~Margoni},
\author[padua]{ M.~Mazzucato}, 
\author[ihep]{V.~Obraztsov},
\author[milano]{M.~Paganoni}, 
\author[ihep]{V.~Poliakov}, 
\author[padua]{F.~Simonetto},
\author[milano]{F.~Terranova}, 
\author[ihep]{E.~Vlasov}. 
              
\address[bologna]{Dipartimento di Fisica, Universit\`a di Bologna and INFN, Bologna, Italy}
\address[cern]{CERN, European Organization for Nuclear Research, Geneva, Switzerland}
\address[lund]{Department of Physics, University of Lund, Lund, Sweden}
\address[milano]{Dipartimento di Fisica, Universit\`a di Milano and INFN, Milan, Italy}
\address[padua]{Dipartimento di Fisica, Universit\`a di Padova and INFN, Padua, Italy}
\address[ihep]{Institute for High Energy Physics, Serpukov, Russia}

\begin{abstract}
A novel technique for longitudinal segmentation of shashlik calorimeters has 
been tested in the CERN West Area beam facility. A 25 tower  very fine samplings
e.m. calorimeter
has been built with vacuum photodiodes inserted in the first 8 radiation 
lengths to sample the initial development of the shower. 
Results concerning  energy resolution, impact point reconstruction and 
$e/\pi$ separation are reported.
\end{abstract}

\end{frontmatter}

\section{Introduction}

In recent years the ``shashlik'' technology has been extensively studied
to assess its performance at $e^+e^-$, $ep$ and $pp$
accelerator experiments \cite{CMS,heraB,LHCB,STIC}. Shashlik calorimeters
 are sampling 
calorimeters in which scintillation light is read-out via wavelength shifting
(WLS) fibers running perpendicularly to the converter/absorber plates 
\cite{fessler,atojan}.
 This technique offers the combination of
an easy  assembly, good hermeticity and fast time response. 
 In many applications it also represents a cheap
solution compared to crystals or cryogenic liquid calorimeters.

Shashlik calorimeters are, in particular, considered  to be good candidates
for barrel 
electromagnetic calorimetry at future linear $e^+e^-$ colliders \cite{TESLA}. 
In this context, the physics requirements impose  
$\sigma(E)/E \leq 0.1/\sqrt{E(GeV)}+0.01$, at least three longitudinal 
samplings, 
transversal segmentation of the order of $0.9^o \times 0.9^o$ 
($\sim 3 \times 3$~cm$^2$) 
and the possibility of performing the read-out in a 3~T magnetic field.
The present shashlik technology can satisfy these requirements, 
except for the optimization of longitudinal segmentation which still needs 
development.
The solution proposed in this paper consists of thin
 vacuum photodiodes inserted
between adjacent towers in the front part of the calorimeter. They
measure the energy deposited in the initial shower development  that allows
for longitudinal sampling and $e/\pi$ separation. A prototype detector 
was exposed to a beam with the aim of measuring the sampling capability 
and demonstrating that the insertion of diodes neither deteriorates critically
the energy response nor produces significant cracks in the tower structure.

\section{The prototype detector}
\label{prototype}

The tested prototype had 25 Pb/scintillator towers,
assembled in a $5\times 5$ matrix.
Each tower consisted  of 140 layers of 1~mm thick lead and
1~mm thick scintillator tiles, resulting in a total depth of $25X_{0}$.
The sampling was the finest ever used with the shashlik technique.
The transversal dimension of each tower was $5 \times 5$~cm$^2$. In the
first $8X_0$ the  tiles had a smaller transverse dimension to provide 
room for the housing of the diodes.
Plastic scintillator consisting of polystyrene doped with 1.5\% paraterphenyl
and 0.05\% POPOP was used. 
Optical insulation between the towers was provided by 
white Tyvek  paper.

As it is custumary in  shashlik technique, the blue light produced in 
the scintillator was carried to the
photodetector at the back of the calorimeter by means of plastic optical 
fibers doped with green WLS.
The 1 mm diameter fibers crossed the tiles in
holes drilled in the lead 
and scintillator plates and they were uniformly distributed  with
a density  of 1 fiber/cm$^2$. In the scintillator tiles the holes
were 2 mm larger (4 mm in the lead) than the fiber diameter.  
 The light transmission between
 the plastic scintillator and the fibers was in air.
 All the fibers from the same
tower were bundled together at the back and connected to photodetectors.
 Two types of fibers were tested:
Bicron BCF20 fibers and Kuraray Y11. In both cases, the emission peak
was at about 500 nm. Light collection was increased by aluminizing
 the fiber end opposite to the photodetector  by
sputtering.

\begin{figure}[htb]
\centerline{\epsfig{file=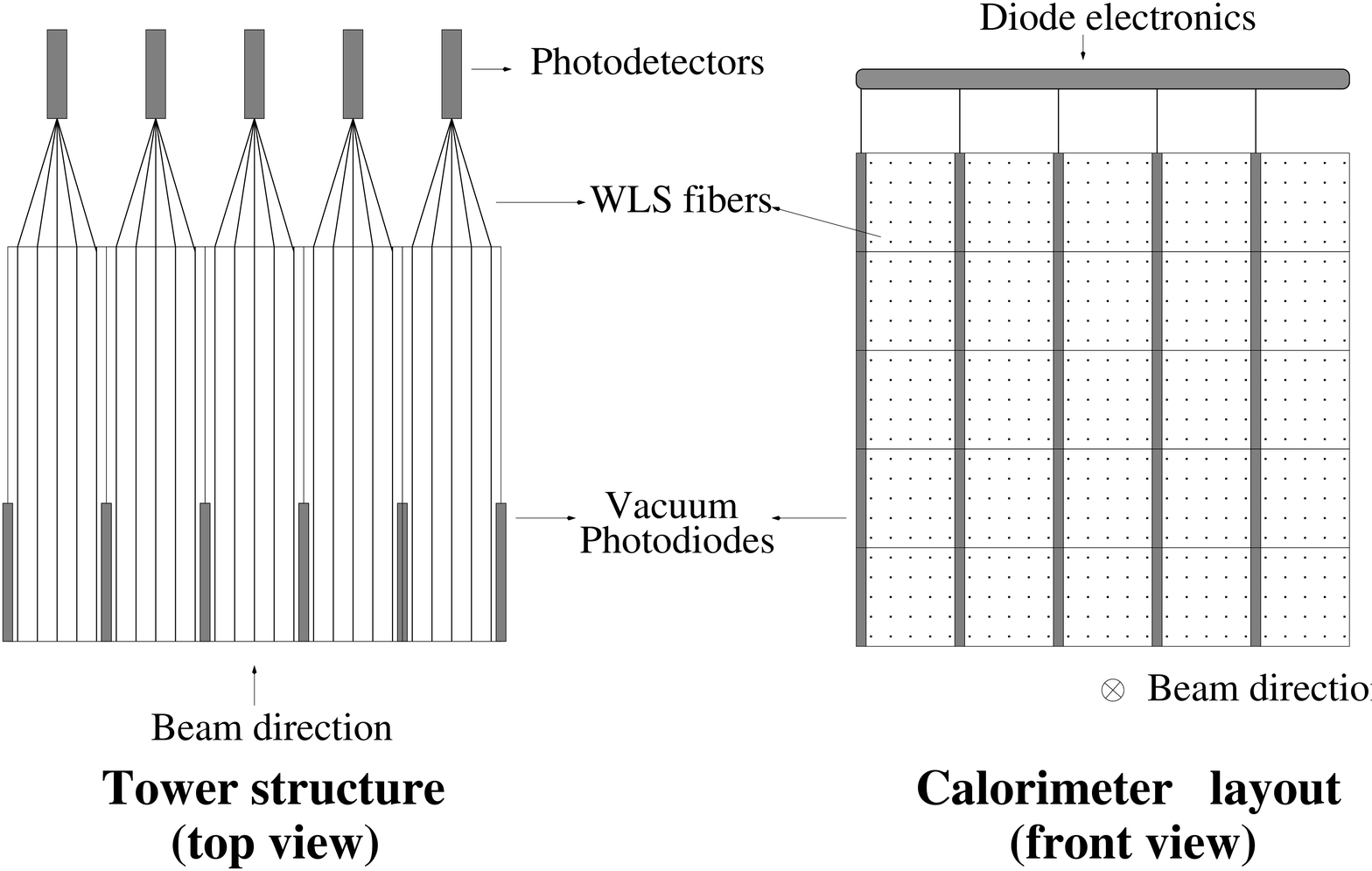,width=7.3cm} }
\caption{Layout of the calorimeter prototype (not in scale).}
\label{install} 
\end{figure}

The light from the fibers was viewed after a 5~mm air gap by 1`` Hamamatsu
R2149-03 phototetrodes. Each tetrode was placed inside an aluminium housing,
containing a charge sensitive JFET preamplifier and a high voltage divider.
The differential output signals
were shaped with a shaping time of 1.500 $\mu$s  and digitized.  
Four towers were read-out with Hamamatsu Avalanche Photodiodes 
instead of tetrodes.
A plexiglass light guide was used to match the smaller APD sensitive 
area to the fiber bundle. Preamplifiers
and voltage dividers were housed in the same mechanical structure as 
the tetrodes.

Two types of vacuum photodiodes, viewed with a bialcali photocathode, 
were produced by EMI \footnotemark[1] (Hamamatsu \footnotemark[2] ) 
with a 
\footnotetext[1]{
EMI vacuum photodiode prototype D437.}
\footnotetext[2]{ Hamamatsu vacuum photodiode prototype SPTXC0046.}
rectangular (squared) front surface of $9 \times 5$ cm$^2$ 
( $5 \times 5$ cm$^2$)
and a  thickness of 5 mm. The diodes were
installed in the first part of the towers 
in order to sample the energy deposited in the first 8~$X_0$.
 They were in optical contact with 
the lateral side of the scintillator tiles and the
light emitted in first part of the detector was
therefore read-out twice since the photons crossing the lateral scintillator
surface were collected by the diode while those reaching the
fibers, either directly or after reflections, were seen by the tetrodes.
Due to the direct coupling, the light collection efficiency
of the diodes was much larger than that of the tetrodes/APD's
and this compensated for the absence of gain in the diodes.

Most of the cells were equipped with EMI vacuum photodiodes.
One diode prototype from Hamamatsu, sampling only
4~$X_0$, was successfully tested during the last part of the data taking. 
Technical characteristics of these devices are listed in table~\ref{diodechar}.
The Hamamatsu prototype dimensions are such that it is possible to house two
diodes in the same tower in order to obtain three longitudinal samplings.
For all diodes,  the same front-end electronics and read-out 
chain as for the tetrodes were used.
The read-out electronics was positioned above the tower stacks
(cfr.fig.\ref{install}).
 
\begin{table}[htb]
\begin{center}
\begin{tabular}{ lll } \hline
  & EMI & Hamamatsu  \\ \hline
Sensitive area                  & 28.9~cm$^2$ & 10.9~cm$^2$ \\
Diode thickness                 & 5.0~mm      & 5.1~mm      \\ 
Working bias                    & -10~V       & -20~V       \\
Capacitance                     & 250~pF      & 17~pF       \\
Energy equivalent e.noise       & $\sim$ 1200 MeV & $\sim$ 900 MeV      \\ \hline
\end{tabular}
\end{center}
\caption{
Technical characteristics of vacuum photodiodes.}
\label{diodechar}
\end{table}

\section{Testbeam setup}

\begin{figure}[htb]
\vspace{1cm}
\centerline{\epsfig{file=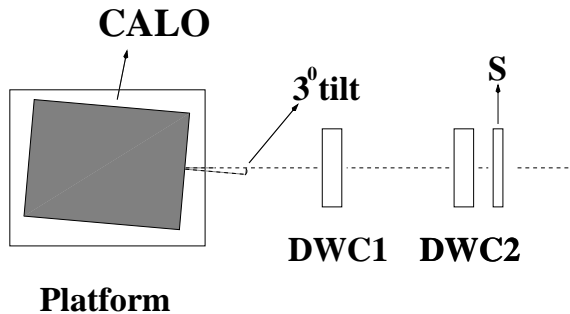,width=7.5cm} }
\caption{Top view of the testbeam setup (not in scale). 
``CALO'' is the calorimeter tilted by 3 degrees with respect to
the beam direction, ``DWC1''-``DWC2'' are the Delay Wire Chambers and 
``S''  the scintillator telescope.
}
\label{testsetup} 
\end{figure}

The prototype was tested
at the X5 beam in the CERN West Area. Electrons ranging from 5 to 75~GeV
and pions of 20, 30 and 50~GeV were used.
The prototype (CALO in fig.\ref{testsetup}) was installed on a moving platform
whose position was controlled at the level of $\simeq 220$~$\mu$m.
In order to avoid particles from channeling through fibers or diodes, 
the calorimeter
was tilted by 3 degrees in the horizontal plane with respect 
to the beam direction.  The absolute 
impact position of the incoming particle was measured by means of two 
Delay Wire Chambers (DWC1 and DWC2)  
with a 2~mm wire pitch and a spatial  resolution of  200~$\mu$m, positioned  
at 0.5 and 1~m  from the calorimeter  frontface. 
External trigger was provided by a layer of scintillators installed near 
DWC2.

A calibration of each tower  was carried out  by exposing the calorimeter
to a  50~GeV 
electron beam  at the beginning of each of the two data taking periods. 
The diode signals were calibrated with  50~GeV 
electrons as well.  
Pedestal runs were taken periodically to monitor
the noise of the electronic amplification chain.

\section{Results}

\subsection{Energy resolution}

\begin{figure}[htb]
\centerline{\epsfig{file=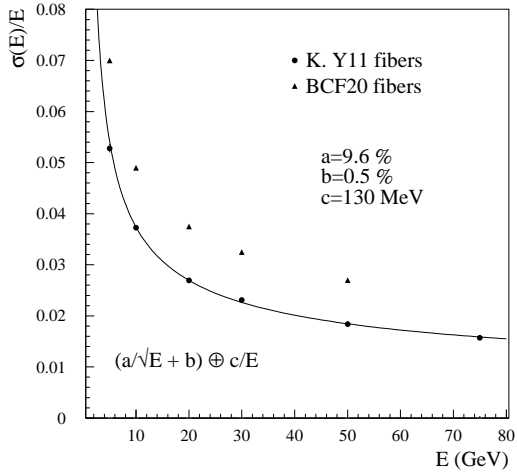,width=7.5cm} }
\caption{Relative energy resolution versus electron beam energy using 
tetrode read-out and KY11 optical fibers. The values obtained using the 
BCF20 fibers are also shown.}
\label{eres} 
\end{figure}

%
%
The energy response is expected to depend on the impact point
since the nearer the fiber the higher
the light collection efficiency. The high fiber density
was used in order  to reduce the non uniformity in light 
response to a level of a few  percent.  
This effect was however not achieved with  BCF20 fibers, due to a 
small scintillating component deteriorating the energy resolution.
KY11 fibers, on the other hand,  had a  non uniformity at the level of
$\pm 1.5\%$. 
\begin{figure}[htb]
\centerline{\epsfig{file=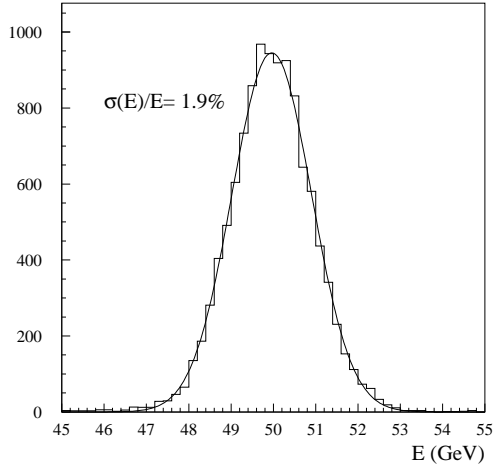,width=7.5cm} }
\caption{Energy resolution for 50~GeV electrons using tetrode  read-out 
and KY11 optical fibers.}
\label{eres_1tow} 
\end{figure}
Fig.\ref{eres_1tow} shows the energy response for 50~GeV electrons in
towers equipped with Kuraray fibers and tetrode redout.
The energy resolution achieved with KY11 fibers 
and tetrode read-out as function of the beam energy  
is shown in fig.\ref{eres} and can be parameterized as\footnotemark[3]
\footnotetext[3]
{Alternatively, by adding the constant term in quadrature: \\
$\frac{\sigma(E)}{E} \ 
= \frac{10.1\%}{\sqrt{E}} \oplus 1.3\%  \oplus \frac{0.130}{E} $ 
}
\begin{center}
\begin{equation}
\frac{\sigma(E)}{E} \ = \ \sqrt{ \left( \frac{9.6\%}{\sqrt{E}}+0.5\% \right)^2 +
\left( \frac{0.130}{E}\right)^2 }
\end{equation}
\end{center}
\noindent where $E$ is expressed in GeV.
The last term corresponds to the electronic noise contribution 
and was measured  from pedestal runs.
A Geant Monte Carlo simulation of the shower development 
in a 1-mm-lead/1-mm-scintillator
sampling gave a smaller value ($\sim 6\%/\sqrt{E}$) for the first term 
of the energy resolution.
Therefore the dominant contribution 
to the measured resolution
was attributed to the photoelectron statistics.

\begin{figure}[htb]
\centerline{\epsfig{file=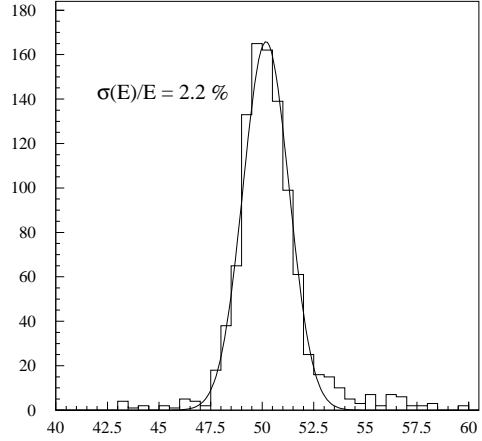,width=7.5cm} }
\caption{Energy resolution for 50~GeV electrons using an APD as photodetector.}
\label{APD} 
\end{figure}

The use of phototetrodes  is not ideal for barrel calorimetry
at $e^+e^-$ colliders. Tetrodes have a rather long longitudinal 
dimension and must be kept at a small angle with respect to the magnetic 
field in order to operate with a maximum gain.
The installation of Avalanche Photodiodes has been proposed
by the CMS collaboration \cite{prop_CMS}
 as an alternative solution. 
Given their very good quantum efficiency ($\sim 80\%$), 
APD should also ensure a better
energy resolution when the photoelectron statistics contribution dominates.
Four APD's were installed in the prototype, as described in section
\ref{prototype}, but unfortunately
no towers were equipped with APD and KY11 fibers.
In fig.\ref{APD} the energy response for 50~GeV electrons impinging on 
a tower equipped with APD is shown.
The high energy tail coming from events reconstructed near the
BCF20 fibers is evident.


\subsection{Linearity}
\begin{figure}[htb]
\centerline{\epsfig{file=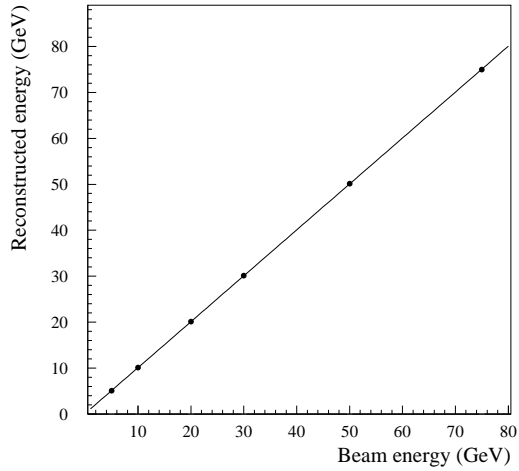,width=7.5cm} }
\caption{Energy reconstructed using the calibration coefficiens computed
 at 50~GeV versus nominal  e$^-$ beam energy.}
\label{linea} 
\end{figure}

Fig. \ref{linea} shows the reconstructed energy versus the nominal electron 
beam
energy when the beam was centered in towers equipped with KY11 fibers. 
No significant deviations from linearity were observed up
to 75 GeV which was the highest energy measured.

\subsection{Spatial resolution}

\begin{figure}[hbt]
\centerline{\epsfig{file=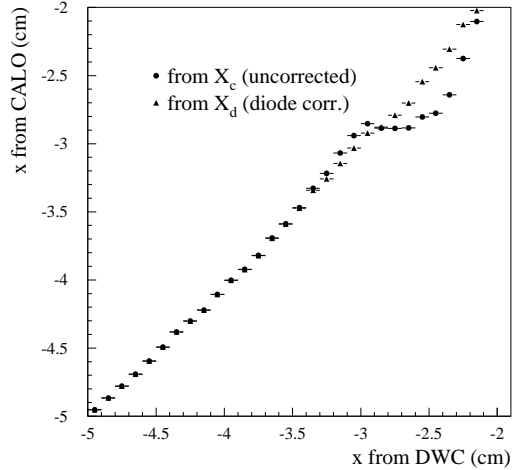,width=7.5cm} }
\caption{Reconstructed horizontal impact position versus beam one (from DWC).
$X_c$ was estimated with eq.(\ref{estim_1}) and $X_d$ with eq.(\ref{estim_2}).}
\label{distortion}
\end{figure}

A position scan along the towers was done using 50~GeV electrons to
establish the precision in the impact point reconstruction. The shower
position reconstruction was based on
%
 center of gravity method
corrected for the detector granularity with the algorithm suggested by
\cite{akopdjanov}. The barycenter
\begin{equation}
X_b= 2 \Delta \sum_{i} i E_{i} / \sum_{i} E_{i} 
\end{equation} 
($\Delta$ is the half-width of the tower and $E_{i}$ the energy deposited 
in tower $i$),
was  modified according to 
\begin{equation}
X_c = b \ \mbox{arcsinh} \left( \frac{X_b}{\Delta} \ \mbox{sinh} \ \delta \right)
\label{estim_1}
\end{equation}
where $b$ is a parameter describing the transversal shower profile
and $\delta \equiv \Delta / b$. 
Since the shower profile was not described by a single exponential,
a two steps procedure was followed: in the first step $X_c'$ was
determined with 
$b=0.85$~cm  and  in the second one the value of $b$ 
was recomputed in the interval $0.45<b<0.85$ according to $X_c'$.
%
$X_c$ was
linear in most of the impact point range, showing non-linearities
only near the diode housing as depicted in fig.\ref{distortion}. 
The non-linear behaviour around the diode  was corrected for by 
using the diode
signal itself.  In particular, in the range of $X_c$ close to the
distortion region, a diode-based estimator was introduced so that
\begin{equation}
X' = X_c + X_d
\label{estim_2}
\end{equation}
where
\begin{equation}
X_d = -b' \log \frac{1}{2} \left( 1+\frac{E_{diode}^{max}}{E^{max+1}} 
\right)+c' 
\ \ ;
\end{equation}
here $E_{diode}^{max}$ is the diode energy in the tower with maximum signal, 
$E^{max+1}$ represents the energy (seen by tetrodes/APD's) in the 
tower closest to the reconstructed impact position and the parameters $b'$ and $c'$ 
were determined with 50~GeV electrons and are $b'=0.2$ cm and $c'=0.3$ cm.

The position resolution of the prototype at the cell center was 1.6~mm
with 50 GeV electrons and had the following energy dependence:
\begin{equation}
\sigma_X(E) \ = \ \sqrt{ \left( \frac{0.9}{\sqrt{E}}\right)^2 +
\left(0.1 \right)^2 }~ \rm{cm}.
\end{equation}

\subsection{Energy leakage to the diode}
\begin{figure}[htb]
\centerline{\epsfig{file=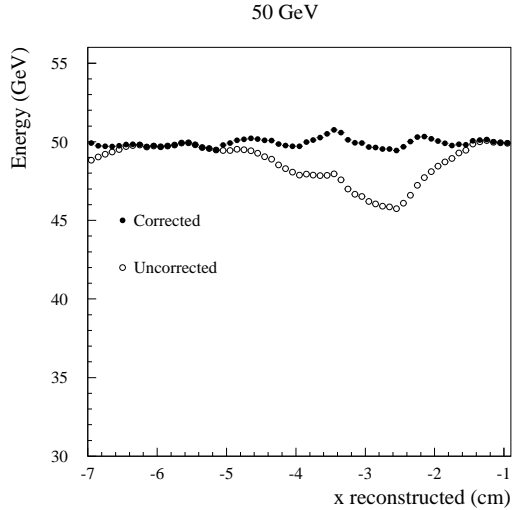,width=7.5cm} }
\caption{Energy versus reconstructed horizontal impact position
before (open circles) and after (black dots) correction.}
\label{crack}
\end{figure}

The dead zone  between two adjacent towers due to the diode
affected only a limited portion of the calorimeter and was always followed by 
a sufficiently long ($>15~X_0$) part of active detector. 
Therefore no complete cracks existed  in the calorimeter. Nevertheless
an energy loss for showers developing near the diode was visible. It was
easily corrected for  by using the reconstructed shower impact point.
The energy response as a  function of the distance $y$
of the reconstructed position  from the two tower border
was parametrized as
\begin{equation}
E(y) = E_0 \cdot (1-~a~ e^{\frac{-y^2}{2 \sigma^2_{\pm}}})
\end{equation}
where $a=0.075$, $\sigma_+ =0.45$ cm for $y > 0$ and 
 $\sigma_- =1.19$ cm for $y < 0$.
Fig.\ref{crack} shows the energy response, before and after the correction,
as function of the reconstructed position for 50 GeV electrons. 
Once the correction was introduced, the remaining   non uniformity
in the energy response was due to  the difference 
in light collection near fibers.

\subsection{Diode response}
\begin{figure}[hbt]
\centerline{\epsfig{file=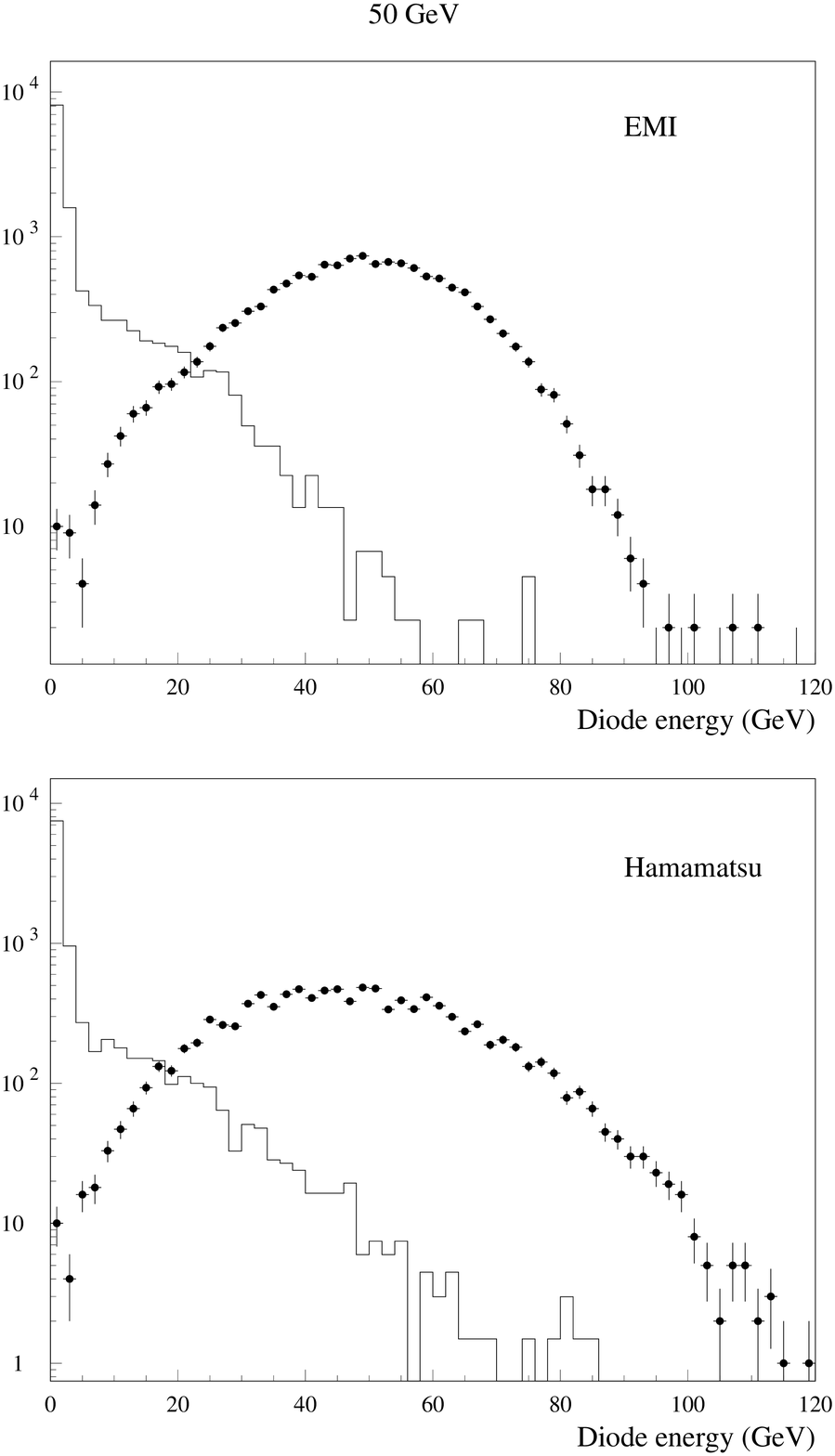,width=7.5cm} }
\caption{Energy response for 50 GeV electrons  (black dots)
and pions (line) for EMI and Hamamatsu diode prototypes.}
\label{diode}
\end{figure}

The EMI and the Hamamatsu diode responses to 50 GeV electrons and pions
are  shown in Fig. \ref{diode}. The widths of  both distributions 
were dominated by the fluctuations in the shower development.
Due to  the different sampling seen  by the two detectors, the light signal
was larger for the EMI and
the fluctuations were more important in the case of the Hamamatsu prototype.
On the other hand the smaller capacitance of the latter ensured 
a much lower electronic noise giving a comparable 
energy equivalent contribution as indicated in 
table~\ref{diodechar}.

\begin{figure}[hbt]
\centerline{\epsfig{file=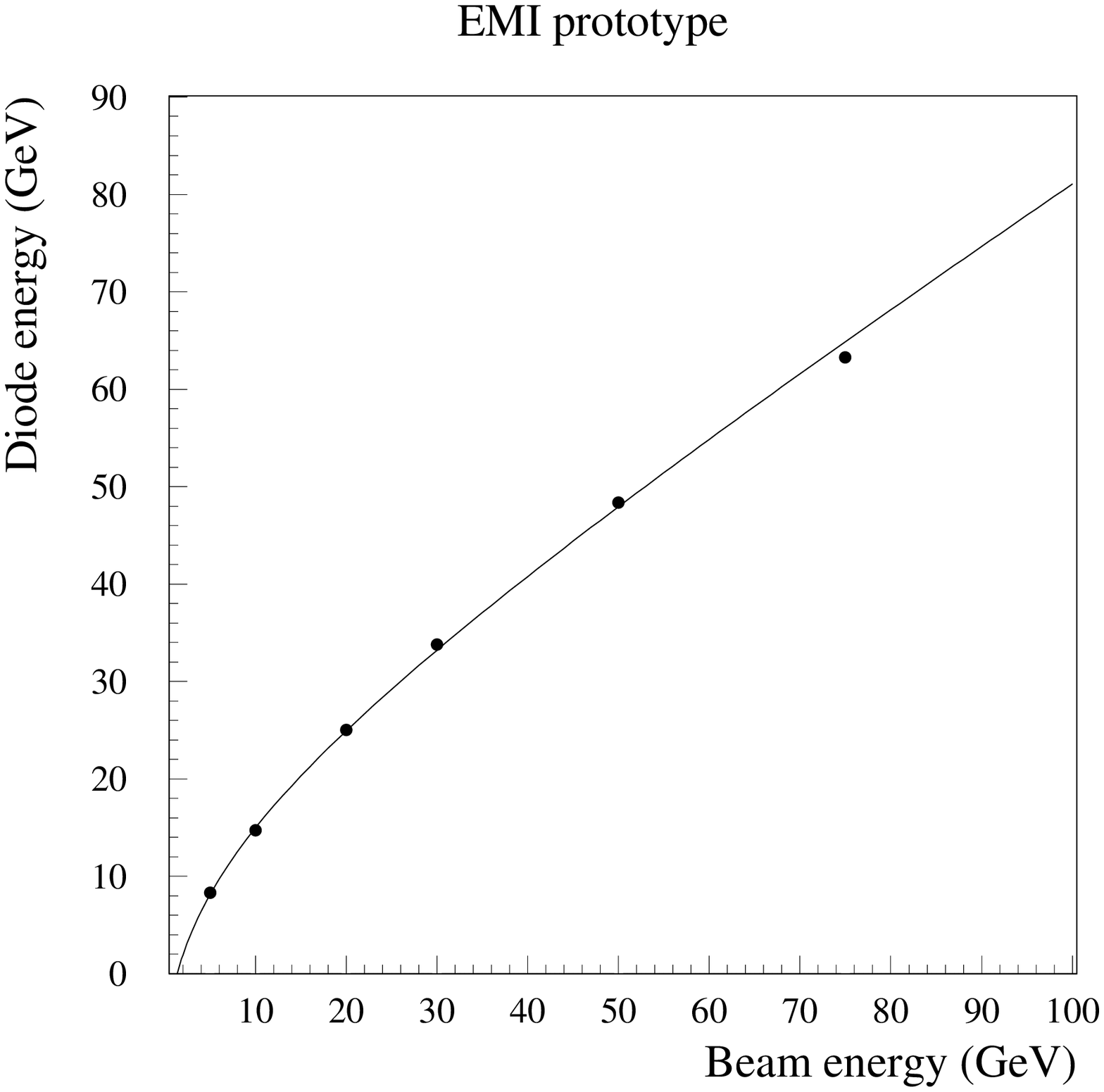,width=7.5cm} }
\caption{Energy response using the calibration coefficients computed at
50 GeV versus nominal electron beam energy (EMI diode).}
\label{lin_diode}
\end{figure}

Since the showers were not contained in the part of the calorimeter read-out by
diodes and the  
longitudinal shower development depends on the energy, the response at
different  electron  energies was not linear as shown in  Fig. \ref{lin_diode}.

\subsection{$e/\pi$ separation}

Separation of electrons from pions was performed using
discriminating variables  based either on purely calorimetric 
data or involving also external information like
the beam energy, known from the settings of main deflection magnet,
 which would be replaced by the momentum estimation 
from the tracking in a collider experiment. The fraction
\begin{equation}
\chi_{E} = \frac{E_{cal}}{E_{beam}}
\end{equation}
can be combined with pure calorimeter variables like the fraction of energy seen by 
the diodes
\begin{equation}
\chi_{D} = \frac{E_{diode}}{E_{cal}}
\end{equation}
and the lateral development of the shower
\begin{equation}
\chi_{S} = \frac{\sum_{i=1}^{N} E_{i} r_i^2 }
               {\sum_{i=1}^{N} E_{i} }
\end{equation}
where  $N$ is the number of towers with signal and $r_i$ the distance of the 
tower from the reconstructed impact position.

\begin{figure}[hbt]
\begin{center}
\begin{tabular}{cc}
\epsfxsize=6.8cm \epsfysize=7.5cm\epsffile{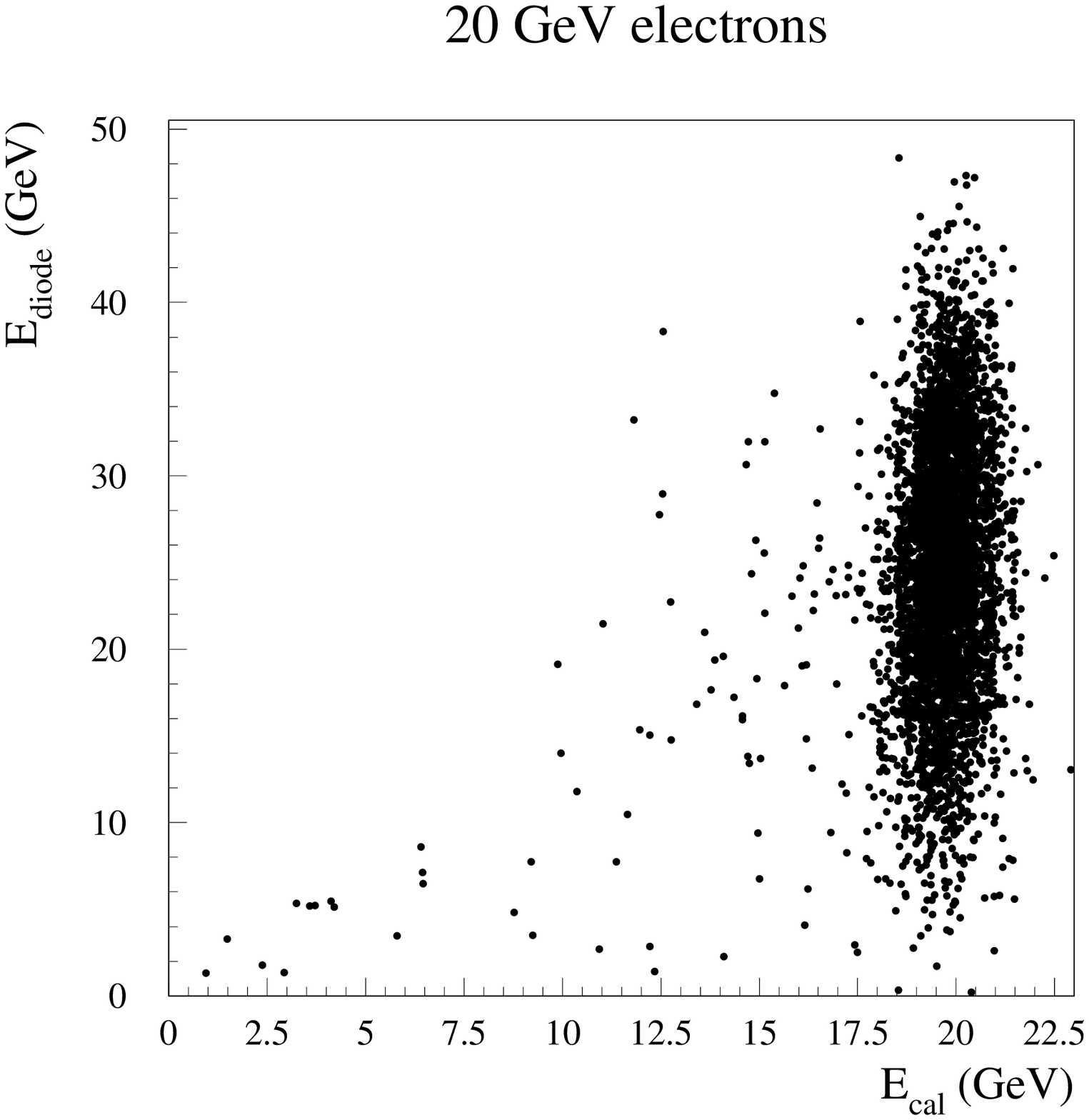}&
\epsfxsize=6.8cm \epsfysize=7.5cm\epsffile{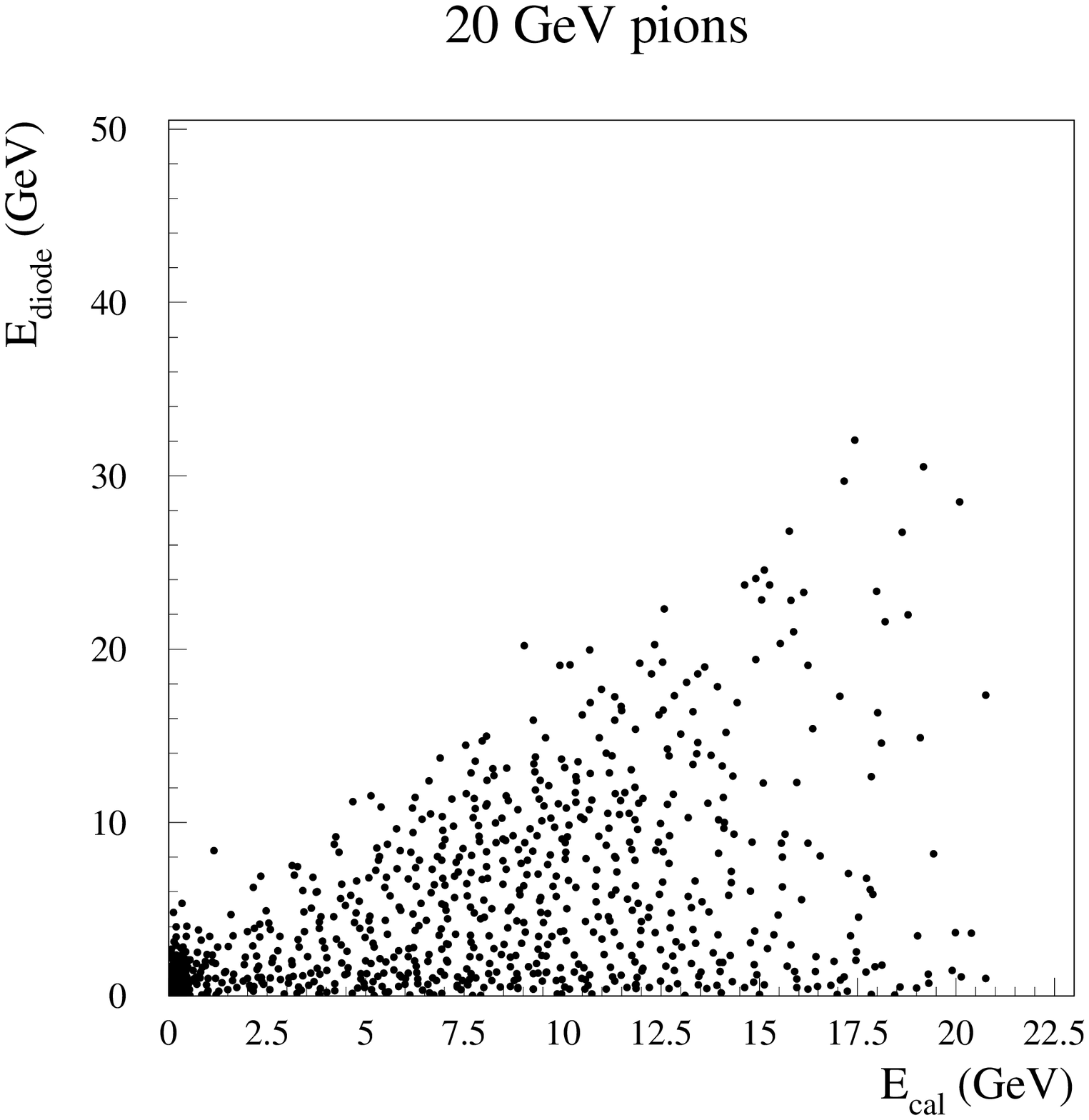}
\end{tabular}
\end{center}
\caption{Diode energy versus total tetrode energy  for $e$ and
$\pi$ at 20~GeV.}
\label{ediode}
\end{figure}

\begin{figure}
\centerline{\epsfig{file=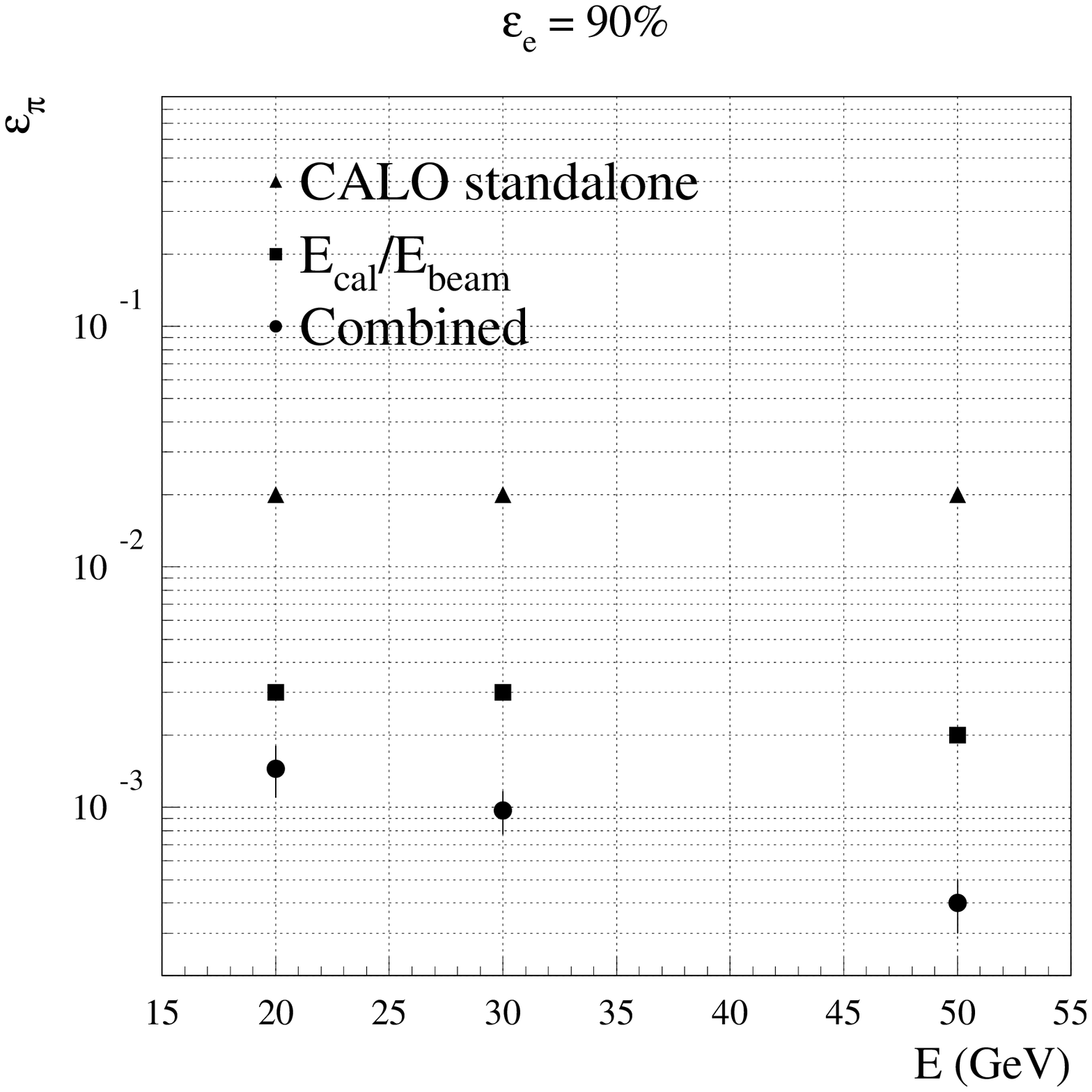,width=7.5cm} }
\caption{Pion contamination versus energy for 90\% electron 
efficiency.}
\label{episep}
\end{figure}

Fig.\ref{ediode} shows $E_{diode}$ versus $E_{cal}$ for pions and electrons
at 20~GeV. The discriminating power of the different variables in terms
of pion contamination for 90\% electron efficiency,
at energies ranging from 20 to 50~GeV is shown in Fig.\ref{episep}. 
In most of the cases, purely calorimetric variables improve the overall
separation capability with a factor $\sim 2$ compared with $\chi_{E}$ by itself.
At 50~GeV the pion contamination for 
 90\% electron efficiency is $(4.0~\pm 1.5)\times 10^{-4}$.

\section{Conclusions}
The present test has demonstrated the technical feasibility of longitudinally 
segmented shashlik calorimeters in which lateral sampling is performed by
vacuum photodiodes. Due to the small  dimension
of the diodes and to the tilt of fibers and diodes with respect to 
the incoming  particles, no significant cracks or dead zones are introduced. 
Performance in terms of energy resolution, impact point reconstruction 
and $e/\pi$ separation seem to be adequate for applications at future  
$e^+e^-$ collider experiments.

\section{Aknowledgements}
 The IHEP workshop staff has 
been essential for the construction of the prototype: we are greatly
indebted with A. Kleschov, P. Korobchuk and A. Tukhtarov.
We wish to thank also C. Fanin for the mechanical project, V. Giordano
and R. Cereseto for the careful work in the realization of the prototype,
G. Rampazzo for the invaluable effort in the construction of 
the read-out chain and
all the staff and technical support of the SL-EA group
for the smooth operation of the accelerator during the testbeam.
A special thank to 
C. Luci for providing the simulation code at the early stage of this work and to
M. Pegoraro and G. Zumerle for usefull suggestions and
for granting the use of EMI diodes.


\begin{thebibliography}{9}
\bibliographystyle{unsrt}
\bibitem{CMS} J.~Badier et al., Nucl. Instr. and Meth. {\bf A348} (1994) 74
\bibitem{heraB} HERA-B Design report, DESY/PRC 95-01.
\bibitem{LHCB} LHCb Technical proposal, CERN/LHCC 98-4.
\bibitem{STIC} S.J.~Alsvaag et al., CERN/EP 98-132.
\bibitem{fessler} H.~Fessler et al., Nucl. Instr. and Meth. {\bf A240}
 (1985) 284.
\bibitem{atojan} G.S.~Atoyan et al., Nucl. Instr. and Meth. {\bf A320}
 (1992) 144.                  
\bibitem{TESLA} R.~Brinkmann, G.~Materlik, J.~Rossbach, A.~Wagner (eds.),
DESY 1997-048.
\bibitem{prop_CMS} CMS Technical Proposal, CERN/LHCC 94-38
\bibitem{akopdjanov} G.A.~Akopdjanov et al., Nucl. Instr. and Meth. {\bf 140}
 (1977) 441.                  
\end{thebibliography}
\end{document}